\documentclass[a4paper,11pt]{article}
\pdfoutput=1
\usepackage{jcappub}
\usepackage[english]{babel}
\usepackage{amsmath,amssymb}
\usepackage{bm}

\renewcommand{\Box}{\square}
\newcommand{\Lcal}{\mathcal{L}}

\newcommand{\tg}{\widetilde g}
\newcommand{\tR}{\widetilde R}
\newcommand{\tX}{\widetilde X}
\newcommand{\MP}{M_*}
\newcommand{\W}{\mathcal{W}}
\newcommand{\Jcal}{\mathcal{J}}
\newcommand{\Ecal}{\mathcal{E}}

\title{AI--Assisted Exploration: DHOST Theories without Quantum Ghosts\footnote{The candidate
transformation studied here was suggested during an exploration with the multi-agent tool
\emph{Denario}~\cite{Denario}. The mathematical statements in the paper are derived independently and
tested with separate symbolic implementations. The AI system is used as a discovery tool, not as a
source of validation.}}

\author[a,b]{Ginevra Braga,}
\affiliation[a]{Gran Sasso Science Institute, Viale F. Crispi 7, I-67100 L'Aquila, Italy}
\affiliation[b]{INFN-Laboratori Nazionali del Gran Sasso, Via G. Acitelli 22, 67100 Assergi (AQ), Italy}
\author[c,d]{Raul Jimenez,}
\affiliation[c]{ICC, University of Barcelona, Mart\'i i Franqu\`es 1, E08028 Barcelona, Spain}
\affiliation[d]{ICREA, Pg. Llu\'is Companys 23, 08010 Barcelona, Spain}
\author[e,f,g,a]{Sabino Matarrese}
\affiliation[e]{Dipartimento di Fisica e Astronomia ``Galileo Galilei'', Universit\`a degli Studi di
Padova, via Marzolo 8, I-35131 Padova, Italy}
\affiliation[f]{INFN, Sezione di Padova, via Marzolo 8, I-35131 Padova, Italy}
\affiliation[g]{INAF-Osservatorio Astronomico di Padova, Vicolo dell'Osservatorio 5, I-35122 Padova,
Italy}
\emailAdd{ginevra.braga@gssi.it}
\emailAdd{raul.jimenez@icc.ub.edu}
\emailAdd{sabino.matarrese@unipd.it}

\abstract{We investigate whether a local symmetry can explain both the classical degeneracy and the
quantum stability of degenerate higher-order scalar--tensor (DHOST) theories. An AI-assisted search
first identifies a simple transformation in a constant-disformal image of Einstein gravity. We show
that this transformation is exactly a field-dependent diffeomorphism plus a vertical shift of a
spectator scalar. This observation leads to a broader result: every regular first-derivative
single-scalar disformal map
$\widetilde g_{\mu\nu}=C(\phi,X)g_{\mu\nu}-D(\phi,X)\nabla_\mu\phi\nabla_\nu\phi$ pulls the spectator
translation back to an exact local gauge redundancy. We derive its closed-form generator and identify
a threefold Jacobian correspondence: the same
$\mathcal J=C-XC_X+X^2D_X$ controls the local inverse of $X\mapsto\widetilde X$, the denominator of the
gauge generator and the determinant of the metric-coordinate field map. This regular construction is
distinct from the known extra symmetry of non-invertible mimetic DHOST. We then prove an obstruction
theorem: adding $K(\phi,\widetilde X)$ for the same scalar preserves the vertical gauge factor if and
only if $K$ is constant. For shift-symmetric nonconstant $K(\widetilde X)$ the local factor is reduced
to the global shift; explicit $\phi$ dependence generally removes even that. The principal-degenerate
cuscuton-type branch remains separate and is not a vertical gauge symmetry. Using regular-orbit
equivalence for invertible derivative-dependent field redefinitions as a lemma, we prove a
regular-orbit spectator theorem: a local DHOST map transports the seed gauge complex, while a local
spectator shift survives precisely when the seed scalar sector is variationally trivial. We extend the explicit
construction to $n$ spectator scalars using the reduced
symmetric-pair Jacobian. The exact multi-field obstruction is variational triviality: for $n<d$ it
forces $K$ to be constant on a generic open jet branch, whereas for $n\geq d$ determinant-type null
Lagrangians give nonconstant topological exceptions. We lift the triangular spectator-ghost coordinate
to an anticanonical transformation and show that the spectator field, its ghost and their dual
antifields form a contractible BV quartet. Hence the full antifield-dependent local BV--BRST
cohomology is isomorphic to that of the seed metric complex: the spectator fibre creates no independent
local counterterm, gauge-deformation or anomaly class. For the pure four-dimensional Einstein seed on
a contractible spacetime, the known Wess--Zumino classification then gives no perturbative local gauge
anomaly; global, topological and boundary questions remain separate. We also formulate a conditional
all-order quantum orbit-equivalence corollary under explicit measure, anomaly and boundary hypotheses.
In the simple constant-map Einstein representative the map fixes the
correlated DHOST coefficient functions and corresponding gauge generator, but the symmetry does not
fix the numerical values of the conformal/disformal constants. For Einstein gravity (or another
two-mode metric seed), the transparent frame has two physical tensor modes for a variationally trivial
scalar sector and two tensors plus one scalar when a regular nonconstant $K$ is added.
The principal original contribution is the explicit regular-disformal spectator construction and its
linked consequences: the closed generator, its Jacobian correspondence, the sharp single- and
multi-field obstructions, and the disformal BV quartet. For
the constant map we also give the exact
Einstein--Hilbert projection, which is quartic Horndeski, and examine the dynamical canonical scalar at
one loop. The known complete local Einstein--scalar divergence contains $(\widetilde\Box\phi)^2$ off shell,
but this term is equation-of-motion redundant at first loop order. After perturbative order reduction,
the essential pole is represented by $\widetilde X^2$; its exact pullback is a first-derivative
$P(\phi,X)$ counterterm throughout the regular disformal orbit. Thus the correlated canonical
one-loop pole introduces no additional perturbative mode at $O(\hbar)$ on the stated regular
hyperbolic EFT domain. This is a controlled one-loop EFT statement, not a calculation for arbitrary
nonlinear $P(X)$, an all-order non-renormalisation theorem or a prediction of a universal massive
ghost.}

\begin{document}
\maketitle

\section{Introduction}
\label{sec:intro}

Scalar--tensor effective field theories (EFTs) provide a general framework for modelling additional
gravitational degrees of freedom. Their derivative expansion inevitably contains higher-derivative
operators. Taken nonperturbatively and without constraints, such operators can introduce an
Ostrogradsky instability~\cite{Ostrogradsky1850,Woodard:2015zca}. DHOST theories avoid the unwanted
mode classically because their highest-derivative kinetic matrix is degenerate: a constraint removes
the extra canonical pair even though the Euler--Lagrange equations are generally higher
order~\cite{Langlois:2017mdk,Langlois:2018dxi,Kobayashi_2019,Langlois_2021}.

Degeneracy and gauge symmetry are logically different. A generic quadratic class-Ia DHOST theory has
two tensor modes and one scalar mode. Degeneracy removes the fourth, Ostrogradsky mode; it does not
usually make the physical scalar gauge. A genuine additional local gauge parameter would normally
remove another canonical pair. The non-invertible mimetic construction is a notable case in which an
extra local metric symmetry is present~\cite{Langlois_2019,Domenech:2023ryc}, but its singular origin
does not establish such a symmetry for regular DHOST theories.

We use two background facts about field redefinitions. A regular first-derivative metric
transformation preserves the constraint structure~\cite{DomenechEtAl:2015DerivativeMetric}, and a general invertible
derivative-dependent transformation gives equivalent Euler--Lagrange systems and transports any seed
Noether identity~\cite{Takahashi:2017zgr}. These are general equivalence statements, not derivations of
the symmetry studied in this paper. Our original problem is to construct the spectator direction
explicitly in the disformal variables, determine when a scalar action obstructs it, and follow its
multi-field, BV and quantum consequences. Singular
branches, where the usual equivalence argument can fail, are kept outside this construction
~\cite{Jirousek:2022jhh}.

The distinction matters quantum mechanically. A gauge identity constrains the counterterm functional,
whereas algebraic degeneracy imposed in one finite operator basis need not be preserved term by term
under renormalisation. At the same time, EFT reasoning warns against diagnosing new particles by
nonperturbatively resumming one loop-suppressed higher-derivative operator. Operators proportional to
the leading equations of motion can be removed order by order by local field redefinitions and do not
add perturbative initial data at that order~\cite{Criado:2018sdb,Glavan:2017jye}. A meaningful quantum
question must therefore distinguish off-shell counterterms, essential on-shell information, finite
matching coefficients, and the domain of the derivative expansion.

In this paper we revisit the question through an AI-assisted symmetry search followed by independent
calculation. The search proposed the constant-map transformation
\begin{equation}
 \delta_\epsilon\phi=(C_0-D_0X)\epsilon,
 \qquad
 \delta_\epsilon g_{\mu\nu}=-2D_0\epsilon\nabla_\mu\nabla_\nu\phi.
 \label{eq:intro-generator}
\end{equation}
Its interpretation produces our first result: modulo ordinary diffeomorphisms,
Eq.~\eqref{eq:intro-generator} is simply a local translation of a spectator coordinate at fixed
disformal metric. We then pull that translation through the general regular first-derivative map
$C(\phi,X),D(\phi,X)$. The result is an exact local gauge generator. Its denominator is the same
kinetic eigenvalue that controls both disformal invertibility and the determinant of the
metric-coordinate field map. Constant or shift-symmetric factors are not necessary; they only make
the formula simpler.

Our second result is a sharp boundary. Once the same scalar enters through
$K(\phi,\widetilde X)$, the vertical shift survives if and only if $K$ is constant. In the
shift-symmetric subcase, a nonconstant $K(\widetilde X)$ breaks the local factor to the global shift.
We prove this off shell, including the explicit-$\phi$ terms in the scalar Euler derivative. This is
an obstruction theorem within a precisely defined regular-disformal mechanism, not a universal
classification of every possible nonlocal, singular, field-enlarged or seed-specific symmetry. The
regular spectator factor itself does admit a complete local BV classification: below we lift the
triangular ghost coordinate to the antifields and prove that the resulting BV quartet is contractible
in the local cohomology modulo the exterior derivative
~\cite{BarnichBrandtHenneaux:1994,BarnichBrandtHenneaux:2000}.

Our regular-orbit spectator theorem combines the two classical results with the standard
Euler--Lagrange chain rule. We prove that, modulo diffeomorphisms, the explicit vertical factor
survives precisely when the seed scalar sector is variationally trivial. This is the general statement
behind the $C(\phi,X),D(\phi,X)$ generator, and it identifies singular mimetic maps and extended-field
constructions as genuinely different mechanisms. The chain rule transports an identity once it is
known; the theorem identifies and classifies the disformal spectator identity itself.

The same argument extends to $n$ scalars. The explicit generator is governed by an $r\times r$
Jacobian on symmetric field-space pairs, with $r=n(n+1)/2$, rather than by a single scalar factor.
This extension makes clear that the regularity boundary is a matrix condition in the multi-field case.
It also exposes a dimension-dependent refinement absent for one scalar: when $n\geq d$, Jacobian-minor
null Lagrangians can be variationally trivial without being constant.

The third result concerns the surviving quantum question. For a canonical scalar seed, the classic
complete Einstein--scalar one-loop divergence is known~\cite{tHooft:1974toh}. Contrary to an
operator-level protection claim, it contains $(\Box\phi)^2$ off shell. We exhibit its exact
equation-of-motion decomposition, retain the essential $X^2$ representative, and pull that correlated
functional through the full regular $C(\phi,X),D(\phi,X)$ orbit. The physical conclusion is positive
but narrower than an exact Ward-identity statement: after $O(\hbar)$ EFT order reduction, the local
pole does not add a fourth perturbative mode on a regular hyperbolic patch.

The original results of this paper form a linked chain. The standard field-redefinition chain rule is
used inside the proof, but is not the result claimed here:
\begin{equation}
 \boxed{\begin{gathered}
  \textbf{explicit regular-disformal spectator generator}\\[-1mm]
  \Downarrow\\[-1mm]
  \text{threefold $\mathcal J$ correspondence and $K(\phi,\widetilde X)$ obstruction}\\[-1mm]
  \Downarrow\\[-1mm]
  \text{regular-orbit spectator criterion and general-$n$ extension}\\[-1mm]
  \Downarrow\\[-1mm]
  \text{topological boundary and explicit disformal BV reduction}
 \end{gathered}}
 \label{eq:original-chain}
\end{equation}
together with the exact EOM-reduced canonical one-loop pullback throughout the same regular orbit and
its conditional equivalence-theorem corollary. We do not claim the disformal map, DHOST degeneracy,
mimetic symmetry, the Einstein--scalar divergence or the equivalence theorem themselves. The metric
Jacobian and its use for invertibility were analysed in Ref.~\cite{Zumalacarregui:2013pma}, which only
noted a possible path-integral role and left that quantum question open. The original content here is
the closed-form spectator generator for the general $C(\phi,X),D(\phi,X)$ map, its exact linkage to all
three appearances of $\mathcal J$, the compact general-$n$ reduced-Jacobian construction, the
dynamical obstruction and its multi-field topological boundary, the explicit anticanonical realization
of the spectator BV quartet, and the orbit-wide canonical one-loop representative. None of the cited
background results derives this connected disformal-spectator construction.

The paper retains the original discovery-to-verification structure. Section~\ref{sec:method} describes
the AI-assisted search and the validation protocol. Section~\ref{sec:classical} introduces quadratic
DHOST and derives the transparent gauge algebra. Section~\ref{sec:general} proves the general
regular-disformal pullback, spells out how the gauge identity selects the restricted constant-map
coefficients, and proves the regular-orbit spectator theorem. Section~\ref{sec:constant} gives the constant-map Einstein--Hilbert
projection. Section~\ref{sec:multifield} gives the explicit multi-field extension. Section~\ref{sec:obstruction} proves the dynamical obstruction and classifies its principal
exceptions. Section~\ref{sec:dof} states the degree-of-freedom and local health conditions.
Section~\ref{sec:quantum} gives the cohomological reduction and the one-loop EFT result. We close by separating the result from mimetic
symmetry and from broader claims about the healthy scalar--tensor landscape.

\section{Methodology: AI-assisted proposal, independent verification}
\label{sec:method}

We used the multi-agent system \emph{Denario}~\cite{Denario} to explore local transformations acting
jointly on $g_{\mu\nu}$ and $\phi$. The tool's role was hypothesis generation. Its proposed
constant-map generator, Eq.~\eqref{eq:intro-generator}, was not assumed to be a symmetry: we retained
the full spacetime dependence of $\epsilon(x)$, varied the defining disformal metric, and computed the
field-space commutator. This exposed a missing ordinary diffeomorphism in a one-generator treatment and
led to the independent Diff-plus-shift basis used below.

The subsequent results were derived without importing AI-produced tensor identities. The
constant-disformal projection was obtained from the connection difference and checked by a separate
component-geometry implementation. The gauge algebra and the BRST differential were tested off shell.
The spectator ghost coordinate was lifted through the BV symplectic potential; the split master action
and the field/ghost plus dual-antifield quartet were then checked in a graded jet algebra and by an
independent lattice cotangent-lift audit, including compatibility of the homotopy with total derivatives.
The one-loop coefficient ledger was reconstructed from the primary Einstein--scalar result, including
the regulator convention and the complete equation-of-motion decomposition. The
$C(\phi,X),D(\phi,X)$ generator, the symmetric-metric determinant, the
$K(\phi,\widetilde X)$ obstruction and the orbit-wide pole density were tested through independent
symbolic routes, including non-diagonal component metrics. All tests are hard failing: a nonzero
residual terminates the suite. Appendix~\ref{app:reproducibility} records the validation scope.

This protocol is important for the scientific conclusion. The automated search found a useful
coordinate expression, but detailed verification changed its interpretation, enlarged its domain from
constant to general regular disformal maps, and ruled out the hoped-for local protection of the
dynamical scalar. The negative part of the result is therefore as central as the discovered symmetry.

\section{Classical theory and transparent gauge structure}
\label{sec:classical}

\subsection{Conventions and quadratic class Ia}

We use metric signature $(-,+,+,+)$ and
\begin{equation}
 X\equiv g^{\mu\nu}\nabla_\mu\phi\nabla_\nu\phi.
 \label{eq:X-convention}
\end{equation}
The conventional k-essence variable often denoted $X_{\rm H}$ is
$X_{\rm H}=-X/2$. The quadratic scalar--tensor Lagrangian is
\begin{equation}
 \Lcal=F R+\sum_{I=1}^{5}A_I\Lcal_I,
 \label{eq:quadratic-action}
\end{equation}
The five quadratic operators are
\begin{align}
 \Lcal_1&=\phi_{\mu\nu}\phi^{\mu\nu},
 &\Lcal_2&=(\Box\phi)^2,\nonumber\\
 \Lcal_3&=(\phi^\mu\phi^\nu\phi_{\mu\nu})\Box\phi,
 &\Lcal_4&=\phi^\mu\phi^\rho\phi_\mu{}^\nu\phi_{\rho\nu},\nonumber\\
 \Lcal_5&=(\phi^\mu\phi^\nu\phi_{\mu\nu})^2,
 &&\phi_{\mu\nu}=\nabla_\mu\nabla_\nu\phi.
 \label{eq:DHOST-basis}
\end{align}
For class Ia, $A_2=-A_1$ and $F-XA_1\neq0$. In the present convention the
remaining degeneracy relations may be written~\cite{BenAchour:2016bwh}
\begin{align}
A_4={\frac{1}{8(F-XA_1)^2}}\Big[&-16XA_1^3+4(3F+16XF_X)A_1^2-X^2FA_3^2
 -(16X^2F_X-12XF)A_3A_1\nonumber\\
&-16F_X(3F+4XF_X)A_1+8F(XF_X-F)A_3+48FF_X^2\Big],
\label{eq:A4-degeneracy}\\
A_5={}&\frac{(4F_X-2A_1+XA_3)
(-2A_1^2-3XA_1A_3+4F_XA_1+4FA_3)}{8(F-XA_1)^2}.
\label{eq:A5-degeneracy}
\end{align}
These conditions remove the Ostrogradsky pair. They do not, by themselves, imply an additional gauge
identity.

\subsection{Constant disformal variables}

Begin with
\begin{equation}
 \tg_{\mu\nu}=C_0g_{\mu\nu}-D_0\phi_\mu\phi_\nu,
 \qquad
 \W=C_0-D_0X,
 \label{eq:constant-map}
\end{equation}
where $C_0,D_0$ are constants and $\phi_\mu=\nabla_\mu\phi$. We work on the
connected Lorentzian branch
\begin{equation}
 C_0>0,
 \qquad \W>0.
 \label{eq:constant-domain}
\end{equation}
The inverse metric, volume element and kinetic variable are
\begin{equation}
 \tg^{\mu\nu}=\frac{1}{C_0}\left(g^{\mu\nu}+\frac{D_0}{\W}\phi^\mu\phi^\nu\right),
 \qquad
 \sqrt{-\tg}=C_0^{3/2}\sqrt{\W}\sqrt{-g},
 \qquad
 \tX=\frac{X}{\W}.
 \label{eq:constant-identities}
\end{equation}

Suppose first that the seed action is a functional of $\tg$ alone. In the transparent coordinates
$(\tg,\phi)$ it possesses ordinary diffeomorphisms and an independent local spectator translation,
\begin{equation}
 \delta_{(\xi,\sigma)}\tg_{\mu\nu}=\mathcal L_\xi\tg_{\mu\nu},
 \qquad
 \delta_{(\xi,\sigma)}\phi=\xi^\rho\phi_\rho+\sigma(x).
 \label{eq:transparent-transformations}
\end{equation}
The vertical part pulls back to
\begin{equation}
 \delta^S_\sigma\phi=\sigma,
 \qquad
 \delta^S_\sigma g_{\mu\nu}=\frac{D_0}{C_0}
 (\phi_\mu\partial_\nu\sigma+\phi_\nu\partial_\mu\sigma),
 \label{eq:constant-vertical}
\end{equation}
and leaves $\tg$ pointwise fixed. Its finite form is
\begin{align}
 \phi'&=\phi+\sigma,\nonumber\\
 g'_{\mu\nu}&=g_{\mu\nu}+\frac{D_0}{C_0}
 \left[\partial_\mu(\phi+\sigma)\partial_\nu(\phi+\sigma)-\phi_\mu\phi_\nu\right].
 \label{eq:finite-vertical}
\end{align}
Consequently $T_{\sigma_2}T_{\sigma_1}=T_{\sigma_1+\sigma_2}$.

\subsection{What the AI-assisted generator is}

For the vector
\begin{equation}
 v_\epsilon^\mu=-D_0\epsilon\phi^\mu,
 \label{eq:v-epsilon}
\end{equation}
an ordinary diffeomorphism gives
\begin{align}
 \mathcal L_{v_\epsilon}g_{\mu\nu}
 &=-2D_0\epsilon\phi_{\mu\nu}
 -D_0(\phi_\mu\partial_\nu\epsilon+\phi_\nu\partial_\mu\epsilon),\nonumber\\
 \mathcal L_{v_\epsilon}\phi&=-D_0X\epsilon.
 \label{eq:diff-pieces}
\end{align}
Adding Eq.~\eqref{eq:constant-vertical} with $\sigma=C_0\epsilon$ cancels the
$\partial\epsilon$ terms. We denote by $E_\epsilon$ the resulting one-parameter transformation---the
AI-assisted candidate reconstructed as this field-dependent diffeomorphism plus vertical shift. It is
given on the fields by
\begin{equation}
 E_\epsilon\phi=\W\epsilon,
 \qquad
 E_\epsilon g_{\mu\nu}=-2D_0\epsilon\phi_{\mu\nu}.
 \label{eq:E-generator}
\end{equation}
Therefore
\begin{equation}
 \boxed{E_\epsilon=\delta^D_{v_\epsilon}+\delta^S_{C_0\epsilon}.}
 \label{eq:E-decomposition}
\end{equation}
Equivalently, $E_\epsilon\tg=\mathcal L_{v_\epsilon}\tg$. The candidate is
neither a pure diffeomorphism nor a new sixth gauge direction: it is a field-dependent basis choice
inside the complete five-parameter Diff-plus-spectator algebra.

For field-independent parameters, Eq.~\eqref{eq:transparent-transformations} has the semidirect-product
bracket
\begin{equation}
 [\delta_1,\delta_2]_{\rm F}=\delta_{(\xi_{12},\sigma_{12})},
 \quad
 \xi_{12}=[\xi_2,\xi_1],
 \quad
 \sigma_{12}=\mathcal L_{\xi_2}\sigma_1-\mathcal L_{\xi_1}\sigma_2,
 \label{eq:semidirect-algebra}
\end{equation}
for the convention $[\delta_1,\delta_2]_{\rm F}=\delta_1\delta_2-\delta_2\delta_1$.
The one-parameter $E_\epsilon$ transformations do not close by themselves for arbitrary local
parameters; their commutator contains an ordinary diffeomorphism. This is why a BRST construction with
only one scalar ghost is incomplete. A minimal independent basis contains the diffeomorphism ghost and
the vertical-shift ghost, as summarised in Appendix~\ref{app:gauge}.

\section{General first-derivative disformal spectator symmetry}
\label{sec:general}

\subsection{Regular map and the threefold Jacobian}

Consider the general first-derivative single-scalar disformal map
\begin{equation}
 \tg_{\mu\nu}=C(\phi,X)g_{\mu\nu}-D(\phi,X)\phi_\mu\phi_\nu.
 \label{eq:general-map}
\end{equation}
Define
\begin{equation}
 \W=C-DX,
 \qquad
 \Jcal=C-XC_X+X^2D_X,
 \qquad
 L=C_\phi-XD_\phi.
 \label{eq:WJ}
\end{equation}
Since $\tX=X/\W$,
\begin{equation}
 \left.\frac{\partial\tX}{\partial X}\right|_\phi=\frac{\Jcal}{\W^2}.
 \label{eq:kinetic-Jacobian}
\end{equation}
Thus the local inverse exists on a branch on which
\begin{equation}
 C\neq0,
 \qquad
 \W\neq0,
 \qquad
 \Jcal\neq0.
 \label{eq:general-domain}
\end{equation}
The last condition is the scalar sector of the usual disformal invertibility criterion
reviewed and derived in Refs.~\cite{Zumalacarregui:2013pma,
DomenechEtAl:2015DerivativeMetric,BenAchour:2016bwh,Takahashi:2017zgr,Babichev:2021bim}.

The same quantity is also the nontrivial factor in the metric-coordinate Jacobian. At fixed $\phi$,
let $h_{\mu\nu}=\delta g_{\mu\nu}$ and define
\begin{equation}
 Q^{(X)}_{\mu\nu}=C_Xg_{\mu\nu}-D_X\phi_\mu\phi_\nu.
 \label{eq:QX-definition}
\end{equation}
The metric block of the Fr\'echet derivative is the rank-one operator
\begin{equation}
 \mathbb A(h)_{\mu\nu}
 =C h_{\mu\nu}-Q^{(X)}_{\mu\nu}
 \phi^\alpha\phi^\beta h_{\alpha\beta}.
 \label{eq:metric-Jacobian-operator}
\end{equation}
On the $N=d(d+1)/2$ dimensional space of symmetric metric components, the matrix determinant lemma
gives
\begin{equation}
 \boxed{\det_{\rm sym}\mathbb A=C^{N-1}\Jcal.}
 \label{eq:metric-Jacobian-determinant}
\end{equation}
In four dimensions this is $C^9\Jcal$. Earlier work analysed the disformal metric Jacobian and its
kinetic eigentensor as tools for frame invertibility~\cite{Zumalacarregui:2013pma,
DomenechEtAl:2015DerivativeMetric};
the possible path-integral role was noted but not developed in Ref.~\cite{Zumalacarregui:2013pma}.
General invertible derivative maps also transport seed gauge generators~\cite{Takahashi:2017zgr}.
Those are inputs. The disformal spectator generator and its exact connection to this Jacobian factor
are derived here.

\subsection{Pullback theorem}

Let $\sigma(x)$ translate $\phi$ at fixed $\tg$. Set
\begin{equation}
 Y=\phi^\rho\nabla_\rho\sigma,
 \qquad
 S_{\mu\nu}=\phi_\mu\nabla_\nu\sigma+\phi_\nu\nabla_\mu\sigma,
 \qquad
 Q^{(\phi)}_{\mu\nu}=C_\phi g_{\mu\nu}-D_\phi\phi_\mu\phi_\nu.
 \label{eq:YQS}
\end{equation}
Varying the map and $X$ gives
\begin{align}
 \delta\tg_{\mu\nu}&=C\delta g_{\mu\nu}+Q^{(X)}_{\mu\nu}\delta X
 +Q^{(\phi)}_{\mu\nu}\sigma-D S_{\mu\nu},
 \label{eq:delta-gtilde-general}\\
 \delta X&=-\phi^\mu\phi^\nu\delta g_{\mu\nu}+2Y.
 \label{eq:delta-X-definition}
\end{align}
Solving $\delta\tg=0$ yields
\begin{equation}
 \boxed{
 \begin{aligned}
 \delta_\sigma\phi&=\sigma(x),\\
 \delta_\sigma X&=\frac{2\W Y+XL\sigma}{\Jcal},\\
 \delta_\sigma g_{\mu\nu}
 &= \frac{D}{C}S_{\mu\nu}
 -\frac{1}{C}Q^{(X)}_{\mu\nu}\delta_\sigma X
 -\frac{1}{C}Q^{(\phi)}_{\mu\nu}\sigma.
 \end{aligned}}
 \label{eq:general-vertical}
\end{equation}
Direct substitution gives the pointwise identity
\begin{equation}
 \boxed{\delta_\sigma\tg_{\mu\nu}=0.}
 \label{eq:gtilde-fixed}
\end{equation}

Equations~\eqref{eq:kinetic-Jacobian}, \eqref{eq:metric-Jacobian-determinant} and
\eqref{eq:general-vertical} establish our first theorem. The papers cited above do not give this closed
local generator for arbitrary regular $C(\phi,X),D(\phi,X)$ or its exact identification with both
Jacobian appearances:

\medskip
\noindent\textbf{Theorem 1 (spectator pullback and Jacobian correspondence).}
\emph{On every branch satisfying Eq.~\eqref{eq:general-domain}, a metric seed
$S_{\rm seed}[\tg]$ written through Eq.~\eqref{eq:general-map} possesses an exact local vertical gauge
redundancy. Its generator is Eq.~\eqref{eq:general-vertical}. The same $\Jcal$ controls the kinetic
inverse, the possible singularity of the generator and the nontrivial factor in the
metric-coordinate determinant.}
\medskip

Physically, this shows that the regularity of the dynamics, the existence of the gauge redundancy, and the equivalence of the field-space description are not independent requirements, but are all governed by the same underlying invertibility condition.
\subsection{Gauge symmetry, disformal data, and degrees of freedom}
\label{sec:factor-selection}

It is useful to distinguish two statements that are sometimes conflated in discussions of disformal
covariance, disformal invariance and the extra symmetry of non-invertible maps
~\cite{Bettoni:2013diz,Langlois_2019,Domenech:2023ryc}. If one chooses a constant map and restricts the original-frame ansatz to the
simple generator~\eqref{eq:E-generator}, cancellation of the terms proportional to
$\partial_\mu\epsilon$ fixes the generator coefficients in terms of the chosen map, not the numerical
values of $C_0,D_0$: the scalar variation is
$\delta_\epsilon\phi=(C_0-D_0X)\epsilon$ and the metric variation is
$\delta_\epsilon g_{\mu\nu}=-2D_0\epsilon\phi_{\mu\nu}$. Pulling back the Einstein--Hilbert seed through
that map then fixes the corresponding DHOST coefficient functions,
\begin{equation}
 F=\mu\sqrt{C_0(C_0-D_0X)},\qquad A_1=2F_X,\qquad
 A_2=-A_1,\qquad A_3=A_4=A_5=0,
 \label{eq:gauge-selection-constant}
\end{equation}
so that the Type-Ia relations hold identically. In this restricted Einstein representative the
transparent seed condition and the gauge identity correlate the coefficient functions: an arbitrary
Type-Ia choice of $F,A_I$ does not have this spectator symmetry. The symmetry alone does not select
particular numerical values of $C_0,D_0$, since all regular choices are field-coordinate charts of the
same type.

The purpose of the regular-orbit theorem, stated in Sec.~\ref{sec:orbit-theorem}, is to separate the
existence of the vertical gauge direction from the particular disformal coordinates used to display
it, and to determine exactly when a same-field scalar sector destroys that direction. The restriction
to constant factors is not a restriction of that theorem. If
the generator is allowed to contain the $\partial_\mu\sigma$, $C_\phi$ and $D_\phi$ terms required by a
derivative-dependent field redefinition, the exact pullback is Eq.~\eqref{eq:general-vertical} for
every $C(\phi,X),D(\phi,X)$ satisfying Eq.~\eqref{eq:general-domain}.  Thus the symmetry fixes the
seed scalar sector (it must be variationally trivial), while the regular functions $C,D$ label the
choice of DHOST coordinates and determine the detailed coefficient functions and the form of the
generator.  The apparent tension between ``constant factors'' and the general result is therefore
only a difference between a restricted generator ansatz and the full conjugated symmetry.

The degree-of-freedom implication is immediate in the transparent variables. For one scalar on a
regular branch, let $N_{\rm seed}$ be the physical mode count of the metric seed. In the absence of
accidental additional constraints

\begin{table}[h!]
    \centering
    \renewcommand{\arraystretch}{1.3}
    \begin{tabular}{c|c|c}
        \hline
        \textbf{Condition} & \textbf{Action} & \textbf{Physical modes} \\
        \hline
        $S=S_{\rm seed}[\tg]+\text{constant}$
        & -- 
        & $N_{\rm seed}$ \\[1mm]
        
        $S=S_{\rm seed}[\tg]+K(\phi,\tX)$,
        $\det(\mathcal K^{\mu\rho}\widetilde g_{\rho\nu})\neq0$
        & Regular map
        & $N_{\rm seed}+1$ \\[1mm]
        
        $C\,\W\,\Jcal=0$
        & Singular-map boundary
        & \text{--} \\
        \hline
    \end{tabular}
    \caption{Selection rule for the number of physical degrees of freedom. 
    A regular spectator sector contributes one additional physical mode, 
    whereas the singular-map boundary corresponds to a change in the 
    propagating-mode structure.}
    \label{tab:dof-selection-rule}
\end{table}

The first line has the seed gauge generators plus the vertical first-class constraint; the second
has lost the vertical constraint and contains one additional regular scalar pair. The last line is the singular
boundary where a change of solution-space structure can occur, but it does not by itself establish a
healthy new mode. For the Einstein--Hilbert seed, $N_{\rm seed}=2$, giving the familiar two- versus
three-mode alternatives. For $n$ spectator scalars, regular kinetic sectors add one scalar mode per
independent nondegenerate sector.
This is the precise sense in which the transparent seed condition and its gauge identity correlate the
DHOST coefficients and determine the spectator contribution to the propagating-mode count.

The finite transformation is most cleanly defined geometrically. If
$F:(g,\phi)\mapsto(\tg,\phi)$ denotes the regular field map and
$t_\sigma:(\tg,\phi)\mapsto(\tg,\phi+\sigma)$, then
\begin{equation}
 T_\sigma=F^{-1}\circ t_\sigma\circ F.
 \label{eq:conjugated-group}
\end{equation}
It follows immediately that the vertical group remains Abelian even though its original-frame
generator is field dependent. Finite transformations are defined while their orbit remains inside the
chosen connected regular patch. For $C_X=D_X=C_\phi=D_\phi=0$,
Eq.~\eqref{eq:general-vertical} reduces to Eq.~\eqref{eq:constant-vertical}. For
$D=0$, $C=C(X)$ and $C_X\neq0$, it instead becomes
\begin{equation}
 \delta_\sigma g_{\mu\nu}=-\frac{2C_X}{C-XC_X}
 g_{\mu\nu}\phi^\rho\nabla_\rho\sigma,
 \label{eq:X-conformal-counterexample}
\end{equation}
which is an explicit regular counterexample to the broad statement that spectator symmetry requires
constant transformation factors.

This symmetry should not be confused with mimetic disformal symmetry. Mimetic DHOST arises on a
non-invertible locus and its familiar extra transformation changes an auxiliary metric while keeping
the mimetic scalar fixed~\cite{Langlois_2019,Domenech:2023ryc}. The present transformation exists on
the opposite, regular branch and translates a spectator scalar at fixed physical metric $\tg$.

\subsection{The regular-orbit spectator theorem}
\label{sec:orbit-theorem}

The proof uses a standard lemma about derivative-dependent field variables. Preservation of
the constraint system has been shown for regular metric transformations
~\cite{DomenechEtAl:2015DerivativeMetric}, and
the transport of pre-existing Noether identities under general invertible derivative-dependent maps
was established in Ref.~\cite{Takahashi:2017zgr}. Neither statement constructs or classifies the
vertical disformal spectator factor. We record the short chain-rule step and then prove that
classification. Let
$q^i$ and $Q^A$
be two local scalar--tensor charts related by an invertible local (possibly derivative-dependent) map
$Q=F[q]$, with a local inverse for its Fr\'echet derivative on the branch under consideration. For a
seed action $S_0[Q]$, define $S[q]=S_0[F(q)]$. We denote the Euler--Lagrange derivatives by
$\Ecal_i[S]\equiv\delta S/\delta q^i$ and
$\Ecal_A[S_0]\equiv\delta S_0/\delta Q^A$, where $i$ and $A$ run over all field components in their
respective charts. Writing $DF$ for the Fr\'echet derivative and $\dagger$ for its formal adjoint, the
Euler--Lagrange chain rule is
\begin{equation}
 \Ecal_i[S]=(DF)^\dagger{}_i{}^A\Ecal_A[S_0].
 \label{eq:EL-chain-rule}
\end{equation}
If $R^A_\sigma[Q]$ is a local seed gauge generator, then
\begin{equation}
 \widehat R^i_\sigma[q]=(DF)^{-1,i}{}_A R^A_\sigma[F(q)]
 \label{eq:orbit-generator}
\end{equation}
obeys $DF\,\widehat R_\sigma=R_\sigma$ and hence
\begin{equation}
 \Ecal_i[S]\widehat R^i_\sigma
 =\Ecal_A[S_0]R^A_\sigma\equiv0.
 \label{eq:orbit-noether}
\end{equation}
Conversely, an identity in the pulled-back chart pushes forward by $DF$ to an identity in the seed
chart. Therefore, on a regular DHOST orbit, an invertible field-coordinate map can obscure a gauge
direction but cannot create a new one or remove one. The local gauge algebra is conjugated by $F$.

We can now state the regular-orbit criterion for the hidden symmetry studied here. Modulo ordinary
diffeomorphisms, a candidate extra generator is the vertical shift
\begin{equation}
 \delta_\sigma\tg_{\mu\nu}=0,
 \qquad
 \delta_\sigma\phi=\sigma(x).
 \label{eq:abstract-vertical}
\end{equation}
For a local scalar sector $L[\tg,\phi]$, Eq.~\eqref{eq:abstract-vertical} is a gauge symmetry if and
only if its scalar Euler derivative is an off-shell identity,
\begin{equation}
 \frac{\delta}{\delta\phi}\int d^dx\sqrt{-\tg}\,L[\tg,\phi]\equiv0.
 \label{eq:variational-triviality}
\end{equation}
Locally in jet space this means that the scalar-dependent part of $L$ is a total divergence, up to
the usual topological cohomology. A genuine scalar kinetic term is therefore incompatible with this
regular hidden gauge factor. For $L=K(\phi,\tX)$, Eq.~\eqref{eq:variational-triviality} is precisely
Theorem~2 below, so the result is independent of the particular $C,D$ representative.

\medskip
\noindent\textbf{Proposition 3 (regular-orbit spectator theorem).}
\emph{For every regular local invertible DHOST field map, a local vertical shift of one scalar exists
on the pulled-back orbit if and only if the seed action is variationally trivial in that scalar
direction. Its Noether identity and algebra are the pullbacks of the corresponding seed identities.
Thus regular
disformal coordinates cannot turn a genuine propagating same-scalar kinetic sector into a gauge mode.
Singular/non-invertible maps, additional compensator fields and nonlocal generators lie outside the
proposition.}
\medskip

The generic transport step is the standard field-redefinition lemma. The proposition's original
content is the regular-disformal spectator classification: the closed vertical generator in
Eq.~\eqref{eq:general-vertical}, the if-and-only-if variational criterion for a same-field scalar
sector, and their Jacobian, multi-field and BV consequences. The regularity qualifier is essential
because Jacobian-singular branches can
evade the ordinary equivalence argument~\cite{Jirousek:2022jhh}.

At the formal quantum level the same statement is a change-of-variables result. If the transformed
measure includes the Berezinian of $DF$, then
\begin{equation}
 \int\!Dq\,\det(DF)\,e^{iS_0[F(q)]}
 =\int\!DQ\,e^{iS_0[Q]}.
 \label{eq:formal-quantum-orbit}
\end{equation}
Consequently a regular field-coordinate map cannot by itself create or remove a physical pole or an
independent Ostrogradsky pair. This is a formal, perturbative statement: a seed counterterm, a
measure/anomaly contribution, an EFT truncation or a singular map can change the quantum conclusion.
The one-loop canonical result below is a concrete order-reduced realization of this orbit-invariant
statement, not an all-order theorem that arbitrary DHOST theories are healthy.

\section{Multi-field regular-disformal extension}
\label{sec:multifield}

The regular-orbit principle is not restricted to one spectator scalar. Multi-disformal maps were
previously studied in cosmological perturbation theory, in singular two-field mimetic constructions, and
more recently in connection with consistent matter couplings
~\cite{Watanabe:2015wja,Firouzjahi:2018xzb,Domenech:2025fermions}. Those studies concern particular
cosmological, singular two-field, or matter-coupling questions. The multi-field map ansatz is therefore
not claimed here. The original result of this section is the regular general-$n$ spectator construction:
the compact reduced
metric determinant and its exact kinetic-Jacobian relation, the pulled-back local $\mathbb R^n$ group,
and the dimension-dependent variational boundary. The cited applications do not construct this local
spectator group or its variational boundary.
Let
$a=1,\ldots,n$ and define the symmetric field-space kinetic matrix
\begin{equation}
 X^{ab}=g^{\mu\nu}\phi^a_\mu\phi^b_\nu,
 \qquad
 \widetilde g_{\mu\nu}=C(\phi,X)g_{\mu\nu}
 -D_{ab}(\phi,X)\phi^a_\mu\phi^b_\nu,
 \qquad D_{ab}=D_{ba}.
 \label{eq:multifield-map}
\end{equation}
Write $r=n(n+1)/2$ for the number of symmetric field-space pairs and
$N=d(d+1)/2$ for the metric components. With matrix multiplication in field space, define
\begin{equation}
 W_a{}^b=C\delta_a{}^b-D_{ac}X^{cb}
 \quad\Longleftrightarrow\quad W=C\mathbf 1-DX,
 \qquad
 \widetilde X=XW^{-1}=(C\mathbf 1-XD)^{-1}X,
 \label{eq:multifield-W}
\end{equation}
where the push-through identity makes the symmetry of $\widetilde X$ manifest and both equalities
follow from the Woodbury formula. The order of $D$ and $X$ matters when they do not commute. A regular
branch requires
$C\neq0$ and $\det W\neq0$ in addition to the reduced metric-map condition below.

For a symmetric pair index $I=(ab)$, define
\begin{equation}
 \mathcal P^I(h)=\phi^{a\mu}\phi^{b\nu}h_{\mu\nu},
 \qquad
 \mathcal Q_{I\,\mu\nu}=C_{,I}g_{\mu\nu}-D_{ab,I}\phi^a_\mu\phi^b_\nu,
 \qquad
 \Jcal_{\rm red}{}^I{}_J=C\delta^I{}_J-\mathcal P^I(\mathcal Q_J),
 \label{eq:multifield-reduced-J}
\end{equation}
with the symmetric-pair convention used consistently in the determinant. The metric Fr\'echet block is
\begin{equation}
 \mathbb A(h)_{\mu\nu}=Ch_{\mu\nu}-\mathcal Q_{I\,\mu\nu}\mathcal P^I(h),
 \label{eq:multifield-A}
\end{equation}
and the matrix-determinant lemma gives
\begin{equation}
 \boxed{\det_{\rm sym}\mathbb A
 =C^{N-r}\det\Jcal_{\rm red}.}
 \label{eq:multifield-metric-det}
\end{equation}
The kinetic Jacobian is an $r\times r$ map on symmetric pairs. Its determinant is related to the
reduced metric determinant by
\begin{equation}
 \boxed{
 \det\!\left(\frac{\partial\widetilde X^I}{\partial X^J}\right)
 =\frac{\det\Jcal_{\rm red}}{[\det W]^{n+1}}.}
 \label{eq:multifield-kinetic-det}
\end{equation}
For $n=1$, $r=1$, $W=\W$ and these equations reduce to
$\det_{\rm sym}\mathbb A=C^{N-1}\Jcal$ and
$\partial\widetilde X/\partial X=\Jcal/\W^2$. The additional $[\det W]^{n+1}$ factor is essential for
$n>1$; the bare kinetic Jacobian is not itself the metric-coordinate determinant.

The pulled-back vertical group is now $\mathbb R^n$. Put
\begin{equation}
 Y^I=\phi^{(a\mu}\nabla_\mu\sigma^{b)},
 \qquad
 S^{ab}_{\mu\nu}=\phi^a_\mu\nabla_\nu\sigma^b+\phi^a_\nu\nabla_\mu\sigma^b,
 \qquad
 \mathcal Q^{(\phi)}_{a\,\mu\nu}=C_{,a}g_{\mu\nu}-D_{bc,a}\phi^b_\mu\phi^c_\nu.
 \label{eq:multifield-sources}
\end{equation}
At fixed $\widetilde g$, the exact local generator is obtained by solving
\begin{equation}
 \boxed{
 \begin{aligned}
 \delta_\sigma\phi^a&=\sigma^a,\\
 \Jcal_{\rm red}{}^I{}_J\,\delta_\sigma X^J
 &=2C Y^I-\mathcal P^I(D_{ab}S^{ab})
   +\mathcal P^I(\mathcal Q^{(\phi)}_a)\sigma^a,\\
 \delta_\sigma g_{\mu\nu}
 &=\frac1C\left[D_{ab}S^{ab}_{\mu\nu}
 -\mathcal Q_{I\,\mu\nu}\delta_\sigma X^I
 -\mathcal Q^{(\phi)}_{a\,\mu\nu}\sigma^a\right].
 \end{aligned}}
 \label{eq:multifield-generator}
\end{equation}
Direct substitution gives $\delta_\sigma\widetilde g_{\mu\nu}=0$. Finite transformations are the
conjugate of $n$ independent translations, so the vertical algebra remains Abelian. The regularity
condition is $\det\Jcal_{\rm red}\neq0$, not a single scalar factor.

For a same-field scalar sector
\begin{equation}
 S_K=\int d^dx\sqrt{-\widetilde g}\,K(\phi^a,\widetilde X^{ab}),
 \label{eq:multifield-K}
\end{equation}
the vertical variation is
\begin{equation}
 \delta_\sigma S_K
 =\int d^dx\sqrt{-\widetilde g}\,\sigma^a
 \left[K_{,a}-2\widetilde\nabla_\mu
   (K_{,\widetilde X^{ab}}\widetilde\phi_b{}^\mu)\right],
 \label{eq:multifield-K-variation}
\end{equation}
up to a boundary term. For $n<d$, at a generic jet one may choose a non-null direction orthogonal to
all $n$ scalar gradients. Varying an independent Hessian along that direction eliminates the terms
containing gradients and forces $K_{,\widetilde X^{ab}}=0$; the Hessian-independent part then gives
$K_{,a}=0$. Continuity extends the result across nongeneric jets within the connected branch. Thus,
for $n<d$, the first-derivative multi-field obstruction reduces to the same constant-$K$ conclusion as
in the single-field theorem.

There is a genuine topological refinement when the number of scalars saturates or exceeds the
spacetime dimension. For $n=d$, on a full-rank Lorentzian branch of fixed orientation,
\begin{equation}
 K_{\rm top}(\phi,\widetilde X)
 =f(\phi)\sqrt{-\det\widetilde X^{ab}}
 \label{eq:multifield-topological-K}
\end{equation}
is nonconstant but variationally trivial, because
\begin{equation}
 \sqrt{-\widetilde g}\,K_{\rm top}
 =f(\phi)\left|\det(\partial_\mu\phi^a)\right|.
 \label{eq:multifield-topological-density}
\end{equation}
On the chosen orientation branch this is the pullback of the target-space top form
$f(\phi)\,d\phi^1\wedge\cdots\wedge d\phi^d$, which is locally exact. For $n>d$, analogous
Jacobian-minor densities arise from closed target-space $d$-forms. These are the familiar
first-derivative null-Lagrangian exceptions in the calculus of variations
~\cite{BallCurrieOlver:1981}; they carry no local scalar dynamics. The exact obstruction in arbitrary
dimension is therefore variational triviality, not literal constancy.

\medskip
\noindent\textbf{Theorem 4 (multi-field regular-disformal spectator theorem).}
\emph{On a branch with $C\neq0$, $\det W\neq0$ and $\det\Jcal_{\rm red}\neq0$, a metric-only seed
has an exact local $\mathbb R^n$ spectator redundancy with generator~\eqref{eq:multifield-generator}.
The reduced Jacobian controls the metric-map singularity and, together with the $W$ factor in
Eq.~\eqref{eq:multifield-kinetic-det}, the kinetic inverse. Adding
$K(\phi^a,\widetilde X^{ab})$ preserves the local group if and only if its scalar density is
variationally trivial. For $n<d$ on a generic open jet branch this requires $K$ to be constant; for
$n\geq d$ determinant/minor null Lagrangians provide nonconstant topological exceptions.}
\medskip

This also sharpens the degree-of-freedom boundary. On the regular branch, an invertible field map
preserves the seed mode count. At $\det\Jcal_{\rm red}=0$ a change of solution-space structure becomes
possible, but the singularity alone does not prove a new healthy degree of freedom; $C=0$, $\det W=0$,
acoustic degeneracy, strong coupling and accidental seed symmetries are separate boundaries. The
single-field locus $\Jcal=0$ is therefore a non-invertible, mimetic-type boundary, not a universal
identification of every mimetic theory~\cite{Jirousek:2022jhh,Gorji:2026dmc}. The null-Lagrangian branch above is different again: it lies
on a regular field-map branch but adds no bulk Euler--Lagrange equation.

\section{Constant-map Einstein projection}
\label{sec:constant}

The constant transformation remains valuable because the Einstein--Hilbert projection and its quantum
Jacobian can be obtained in closed form. Let
\begin{equation}
 S_{\rm EH}[\tg]=\frac{\MP^2}{2}\int d^4x\sqrt{-\tg}\,\tR,
 \qquad \mu\equiv\frac{\MP^2}{2}.
 \label{eq:EH-seed}
\end{equation}
The connection difference is
\begin{equation}
 \widetilde\Gamma^\lambda{}_{\mu\nu}-\Gamma^\lambda{}_{\mu\nu}
 =-\frac{D_0}{\W}\phi^\lambda\phi_{\mu\nu}.
 \label{eq:connection-difference}
\end{equation}
After contracting the curvature and reducing one Ricci-tensor term by parts, the exact action is
Eq.~\eqref{eq:quadratic-action} with
\begin{equation}
 \boxed{
 \begin{aligned}
 F(X)&=\mu\sqrt{C_0(C_0-D_0X)},\\
 A_1(X)&=-\mu D_0\sqrt{\frac{C_0}{C_0-D_0X}}=2F_X,\\
 A_2&=-A_1,\\
 A_3&=A_4=A_5=0.
 \end{aligned}}
 \label{eq:projected-coefficients}
\end{equation}
Moreover
\begin{equation}
 F-XA_1=\mu\frac{C_0^{3/2}}{\sqrt\W}\neq0
 \label{eq:FminusXA1}
\end{equation}
on the regular branch. This DHOST-chart denominator is not numerically identical to the field-map
factor $\Jcal$: for the constant map, $\Jcal=C_0$ whereas Eq.~\eqref{eq:FminusXA1} gives the
additional factor $\mu C_0^{3/2}/\sqrt{\W}$. Their simultaneous nonvanishing is a regularity
statement, not an equality of the two quantities. Substitution into Eqs.~\eqref{eq:A4-degeneracy} and
\eqref{eq:A5-degeneracy} returns $A_4=A_5=0$ identically.

The classification is important. With $X_{\rm H}=-X/2$ and
$G_4(X_{\rm H})=F(-2X_{\rm H})$, one has $G_{4,X_{\rm H}}=-2F_X$. Therefore
\begin{align}
 \Lcal_4^{\rm H}
 &=G_4R+G_{4,X_{\rm H}}
 \left[(\Box\phi)^2-\phi_{\mu\nu}\phi^{\mu\nu}\right]\nonumber\\
 &=FR+2F_X\left[\phi_{\mu\nu}\phi^{\mu\nu}-(\Box\phi)^2\right].
 \label{eq:Horndeski-translation}
\end{align}
The projected theory is quartic Horndeski, not strictly beyond Horndeski. This agrees with the
preservation of the Horndeski form under constant disformal transformations
~\cite{Bettoni:2013diz}. General regular $C(\phi,X),D(\phi,X)$ maps reach a wider quadratic-DHOST orbit
~\cite{BenAchour:2016bwh}; the constant projection is a controlled worked benchmark, not the boundary
of the spectator theorem and not the only healthy scalar--tensor theory.

The operator label ``quartic Horndeski'' does not by itself fix the mode count. A generic quartic
Horndeski theory has only the four diffeomorphism gauge directions and therefore propagates three
modes. The special functions in Eq.~\eqref{eq:projected-coefficients}, however, are not a generic
choice: they are the invertible image of Einstein--Hilbert gravity with a spectator scalar and inherit
the additional vertical first-class generator. Counting the six spatial-metric components and the
scalar gives, respectively,
\begin{equation}
 N_{\rm dof}^{\rm projected}
 =\frac12\!\left[2(6+1)-2(4+1)\right]=2,
 \qquad
 N_{\rm dof}^{\rm generic\ H4}
 =\frac12\!\left[2(6+1)-2(4)\right]=3.
 \label{eq:constant-Horndeski-dof-count}
\end{equation}
Thus this is a two-mode, gauge-redundant subclass written in quartic-Horndeski form; adding a regular
nonconstant $K(\phi,\widetilde X)$ removes the vertical generator and restores the usual third scalar
mode.

\section{Dynamical obstruction for \texorpdfstring{$K(\phi,\widetilde X)$}{K(phi,X)}}
\label{sec:obstruction}

\subsection{Off-shell Noether identity}

Promote the same scalar by considering
\begin{equation}
 S[\tg,\phi]=\int d^dx\sqrt{-\tg}
 \left[\frac{\MP^2}{2}\tR+K(\phi,\tX)\right],
 \qquad
 \tX=\tg^{\mu\nu}\phi_\mu\phi_\nu.
 \label{eq:dynamical-seed}
\end{equation}
Ordinary diffeomorphisms survive. The vertical variation at fixed $\tg$ is
\begin{align}
 \delta_\sigma S_K
 &=\int d^dx\sqrt{-\tg}\left[
 K_\phi\sigma+2K_{\tX}\widetilde\phi^\mu
 \widetilde\nabla_\mu\sigma\right]\nonumber\\
 &=\int d^dx\sqrt{-\tg}\,\sigma\left[
 K_\phi-2\widetilde\nabla_\mu(K_{\tX}\widetilde\phi^\mu)\right],
 \label{eq:K-variation}
\end{align}
up to a boundary term. An arbitrary local $\sigma(x)$ would require the off-shell identity
\begin{equation}
 K_\phi-2\widetilde\nabla_\mu(K_{\tX}\widetilde\phi^\mu)\equiv0.
 \label{eq:vertical-Noether-condition}
\end{equation}
Expanding it at a point gives
\begin{equation}
 K_\phi-2\tX K_{\phi\tX}
 -2\left(K_{\tX}\tg^{\mu\nu}
 +2K_{\tX\tX}\widetilde\phi^\mu\widetilde\phi^\nu\right)
 \widetilde\phi_{\mu\nu}\equiv0.
 \label{eq:jet-identity}
\end{equation}
The symmetric Hessian is an arbitrary off-shell jet. A component transverse to a non-null scalar
gradient enforces $K_{\tX}=0$ on an open kinetic branch. It follows that
$K_{\phi\tX}=K_{\tX\tX}=0$, and the Hessian-independent part then enforces $K_\phi=0$.
Continuity extends the conclusion through zero-gradient points on the connected branch.

\medskip
\noindent\textbf{Theorem 2 (dynamical obstruction).}
\emph{For $d\ge2$, the regular-disformal vertical symmetry of Theorem~1 survives the addition of
$K(\phi,\widetilde X)$ for the same scalar if and only if $K$ is constant on the connected branch.
For a nonconstant shift-symmetric $K(\widetilde X)$, the local factor is broken to the global shift
$\phi\mapsto\phi+\mathrm{constant}$; with explicit $\phi$ dependence, that global shift is generally
absent as well.}
\medskip

The proof is independent of $C,D$ and all of their derivatives because it is performed at fixed
transparent metric.
Accordingly it applies throughout the regular orbit in Eq.~\eqref{eq:general-domain}. It does not cover
singular maps, extra compensator fields, nonlocal transformations, or transformations proportional to
the equations of motion.

\subsection{Principal determinant and the cuscuton boundary}

The scalar principal tensor is
\begin{equation}
 \mathcal K^{\mu\nu}=2K_{\tX}\tg^{\mu\nu}
 +4K_{\tX\tX}\widetilde\phi^\mu\widetilde\phi^\nu.
 \label{eq:acoustic-tensor}
\end{equation}
The matrix determinant lemma yields in $d$ dimensions
\begin{equation}
 \det(\mathcal K^{\mu\rho}\tg_{\rho\nu})
 =2^dK_{\tX}^{d-1}(K_{\tX}+2\tX{}K_{\tX\tX}).
 \label{eq:acoustic-determinant}
\end{equation}
An identically degenerate principal symbol on an open kinetic interval has two branches. The first,
$K_{\tX}=0$, has no kinetic principal part; it is the spectator theory only when $K_\phi=0$, while a
nonconstant $K(\phi)$ is auxiliary rather than gauge. The second solves
\begin{equation}
 K_{\tX}+2\tX{}K_{\tX\tX}=0,
 \qquad
 K(\phi,\tX)=K_0(\phi)+K_1(\phi)\sqrt{|\tX|}
 \label{eq:cuscuton-branch}
\end{equation}
on a connected sign branch, where the integration functions may depend on $\phi$. This is the
cuscuton-type kinetic structure
~\cite{Afshordi:2006ad}. It is a constrained/nonpropagating scalar limit, not a regular propagating
k-essence mode, and it does not satisfy Eq.~\eqref{eq:vertical-Noether-condition} when $K_1\neq0$.
Principal degeneracy is therefore not equivalent to vertical gauge symmetry. An isolated zero of
either determinant factor is a background kinetic or strong-coupling boundary, not a new off-shell
gauge identity.

\section{Degrees of freedom and local health}
\label{sec:dof}

The degree-of-freedom count is transparent in $(\tg,\phi)$. For the Einstein--Hilbert seed used in the
explicit representative, constant $K$ leaves the scalar absent from the Hamiltonian and its momentum
is the first-class constraint $\pi_\phi\approx0$. Together with the four diffeomorphism generators, the
seven configuration variables $(\widetilde h_{ij},\phi)$ have five first-class gauge directions and
hence two physical modes. A nonconstant $K(\phi)$ with no kinetic
term instead makes the scalar auxiliary and does not possess this vertical gauge factor. For a regular,
nondegenerate kinetic $K(\phi,\tX)$, the vertical constraint is lost while the four diffeomorphism
generators remain, leaving three physical modes. This is an equivalence-frame count; we do not present
it as a new original-frame Dirac derivation of general DHOST degeneracy.

On the regular branch, invertibility preserves the constraint algebra and local solution
space~\cite{DomenechEtAl:2015DerivativeMetric,Takahashi:2017zgr}, but it does not make every
background healthy. For a timelike scalar gradient in our mostly-plus convention, the scalar has the
canonical sign and a hyperbolic acoustic cone when~\cite{GarrigaMukhanov:1999,
ArmendarizPicon:2000ah}
\begin{equation}
 K_{\tX}<0,
 \qquad
 K_{\tX}+2\tX{}K_{\tX\tX}<0,
 \label{eq:kessence-health}
\end{equation}
with sound speed
\begin{equation}
 c_s^2=\frac{K_{\tX}}{K_{\tX}+2\tX{}K_{\tX\tX}}.
 \label{eq:sound-speed-general}
\end{equation}
The sound speed is not used to count degrees of freedom or to prove the symmetry. We display it because
a three-mode branch is physically admissible only where its scalar equation is hyperbolic and free of
gradient instabilities; it also defines the local hyperbolic domain assumed in the one-loop EFT
statement below.
One must additionally impose the regular disformal branch, weak coupling, background-variation scales
below the cutoff, and compatible matter couplings. ``Degenerate'' is therefore a statement about the
Ostrogradsky pair, not a synonym for positive energy, hyperbolicity, or phenomenological viability.

\section{One-loop EFT benchmark}
\label{sec:quantum}

\subsection{The \texorpdfstring{$K=0$}{K=0} spectator path integral and measure}

For the general map, the field-coordinate Fr\'echet derivative from $(g,\phi)$ to $(\tg,\phi)$ is block
triangular: its metric diagonal block is $\mathbb A$ in
Eq.~\eqref{eq:metric-Jacobian-operator}, its scalar diagonal block is the identity, and derivatives
acting on $\delta\phi$ occur only in the off-diagonal block. Hence its determinant is
$C^{N-1}\Jcal$ in the single-field case (and $C^{N-r}\det\Jcal_{\rm red}$ for the multi-field
metric block). For nonconstant $C,D$ this is a field-dependent but ultralocal measure factor, formally
proportional to $\delta^{(d)}(0)\int d^dx\ln|C^{N-1}\Jcal|$. The standard dimensional-regularisation
prescription sets this scaleless contact term to zero; in a different scheme the corresponding local
measure term must be carried explicitly~\cite{Zumalacarregui:2013pma,Criado:2018sdb}. In either case it
depends only on $\phi,X$ and does not create a higher-derivative kinetic operator.

Choosing the vertical gauge $\phi=0$ gives an algebraic Faddeev--Popov block equal to the identity. On
that slice $X=0$ and $\det_{\rm sym}\mathbb A=C(0,0)^N$, a field-independent normalization. Formally,
on a connected regular patch and with the transformed measure treated consistently,
\begin{equation}
 Z_{K=0}=\mathcal N\!\left(C(0,0)\right)Z_{\rm metric}[\tg].
 \label{eq:K0-partition}
\end{equation}
Thus counterterms respecting the classical gauge complex are correlated pullbacks of functionals of
$\tg$ that are invariant under diffeomorphisms, modulo BRST-exact terms and total derivatives; when
classifying physical EFT operators one may additionally quotient by leading equation-of-motion terms. This formal
reduction assumes compatible measures, boundary conditions, gauge fixing, and treatment of zero modes;
it is not by itself an all-order anomaly theorem.

\subsection{Full local BV cohomology of the regular spectator fibre}
\label{sec:spectator-cohomology}

The spectator sector admits a sharper local statement. In the transparent variables, use the BRST
rules in Eq.~\eqref{eq:BRST}. The standard contractible-pair construction motivates a triangular
change of ghost coordinates~\cite{BarnichBrandtHenneaux:1994,BarnichBrandtHenneaux:2000}; its specific
disformal-spectator realization here is
\begin{equation}
 \widehat\rho\equiv c^\mu\partial_\mu\phi+\rho=s\phi,
 \qquad
 \rho=\widehat\rho-c^\mu\partial_\mu\phi.
 \label{eq:rhohat}
\end{equation}
Off-shell nilpotency gives
\begin{equation}
 s\phi=\widehat\rho,\qquad s\widehat\rho=0,
 \label{eq:contractible-pair}
\end{equation}
while the transformations of $\tg_{\mu\nu}$ and $c^\mu$ contain neither member of this pair. The same
holds for every prolonged jet,
$s(\partial_I\phi)=\partial_I\widehat\rho$ and
$s(\partial_I\widehat\rho)=0$. With
\begin{equation}
 \kappa=\sum_I(\partial_I\phi)\frac{\partial_l}{\partial(\partial_I\widehat\rho)},
 \qquad
 N_{\rm pair}=\sum_I\left[
 (\partial_I\phi)\frac{\partial}{\partial(\partial_I\phi)}
 +(\partial_I\widehat\rho)\frac{\partial_l}{\partial(\partial_I\widehat\rho)}
 \right],
 \label{eq:contracting-homotopy}
\end{equation}
one has $\{s_{\rm pair},\kappa\}=N_{\rm pair}$. The prolonged homotopy commutes with total
derivatives, so the standard contractible-pair argument applies also modulo the exterior derivative.

The restriction to fields and ghosts is unnecessary. The triangular coordinate
Eq.~\eqref{eq:rhohat} has the local anticanonical cotangent lift
\begin{equation}
 \widehat\rho^*=\rho^*,\qquad
 \widehat\phi^*=\phi^*+\partial_\mu(\rho^*c^\mu),\qquad
 \widehat c^*_{\mu}=c^*_{\mu}-\rho^*\partial_\mu\phi,
 \label{eq:BV-antifield-lift}
\end{equation}
with the metric variables and antifields unchanged. The antifields are densities. Under the same
boundary hypothesis as the master equation, anticanonicity follows directly from
\begin{align}
 \phi^*\delta\phi+\rho^*\delta\rho+c^*_{\mu}\delta c^\mu
 ={}&\widehat\phi^*\delta\phi+\widehat\rho^*\delta\widehat\rho
 +\widehat c^*_{\mu}\delta c^\mu\nonumber\\
 &-\partial_\mu\!\left(\rho^*c^\mu\delta\phi\right).
 \label{eq:BV-symplectic-potential}
\end{align}
Substituting Eq.~\eqref{eq:BV-antifield-lift} and
$\rho=\widehat\rho-c^\mu\partial_\mu\phi$ into Eq.~\eqref{eq:BV-master} gives, modulo the same
surface term,
\begin{equation}
 S_{\rm min}=S_{\rm BV}^{\rm seed}[\tg,c;\tg^*,\widehat c^*]
 +\int d^dx\,\widehat\phi^*\widehat\rho.
 \label{eq:BV-split-master}
\end{equation}
Thus the full spectator differential is
\begin{equation}
 s\phi=\widehat\rho,\qquad s\widehat\rho=0,
 \qquad s\widehat\rho^*=\widehat\phi^*,\qquad s\widehat\phi^*=0.
 \label{eq:BV-quartet}
\end{equation}
Besides the field--ghost pair there is therefore a dual antifield pair. For their prolonged jets the
contracting homotopy is
\begin{equation}
 \kappa_{\rm BV}=\sum_I\left[
 (\partial_I\phi)\frac{\partial_l}{\partial(\partial_I\widehat\rho)}
 +(\partial_I\widehat\rho^*)
 \frac{\partial_l}{\partial(\partial_I\widehat\phi^*)}\right],
 \qquad \{s_{\rm quartet},\kappa_{\rm BV}\}=N_{\rm quartet}.
 \label{eq:BV-quartet-homotopy}
\end{equation}
It also commutes with horizontal total derivatives. The standard contractible-pair theorem therefore
removes the entire quartet from the local cohomology, including its Koszul--Tate/antifield sector
~\cite{BarnichBrandtHenneaux:1994,BarnichBrandtHenneaux:2000}.

The abstract contractible-pair theorem is the input. Our original step is its explicit local
anticanonical realization for the disformal spectator fibre: Eqs.~\eqref{eq:BV-antifield-lift} and
\eqref{eq:BV-split-master} exhibit the transformed ghost and both antifield partners and reduce the
complete master action, rather than only its field--ghost truncation.

\medskip
\noindent\textbf{Theorem 5 (full local BV spectator-cohomology reduction).}
\emph{On a connected regular $K=0$ patch, under the standard local jet-algebra and boundary
hypotheses and the seed properness assumption of Appendix~\ref{app:gauge}, the full local BV--BRST
cohomology is, for every ghost number $g$ and form degree $p$,
\begin{equation}
 \boxed{
 H^{g,p}_{\rm loc}(s\mid d;\tg,c,\tg^*,c^*,\phi,\rho,\phi^*,\rho^*)
 \simeq H^{g,p}_{\rm loc}(s_{\rm seed}\mid d;\tg,c,\tg^*,c^*).}
 \label{eq:local-cohomology-reduction}
\end{equation}
For $n$ spectators the same statement follows from the $n$ independent BV quartets generated by
$\widehat\rho^a=s\phi^a$.}
\medskip

An invertible local disformal map with a local inverse has an anticanonical cotangent lift generated by
its Fr\'echet derivative, so Eq.~\eqref{eq:local-cohomology-reduction} is transported unchanged from
the transparent variables to every regular chart in the orbit. This step fails on the singular locus.

At ghost number zero, including antifields, the spectator fibre therefore creates no independent
local counterterm, consistent action deformation, or deformation of the gauge transformations or
their algebra: every surviving class is inherited from the seed BV complex. At ghost number one it
creates no independent local anomaly class. This is a complete local statement about the regular
spectator factor, not a classification of additional gauge symmetries or anomaly classes already
present in an unspecified seed. It also does not cover nonlocal observables, global topology, physical
boundaries without a BV--BFV completion, or singular field-map surfaces. The dynamical $K\neq0$
theory has no vertical ghost and lies outside the quartet result.

For the pure Einstein seed the remaining local anomaly question is known. On a four-dimensional
spacetime of topology $\mathbb R^4$, the full Wess--Zumino analysis gives
$H^{1,4}_{\rm loc}(s_{\rm Diff}\mid d)=0$ and shows that antifield-dependent solutions are
cohomologically trivial~\cite{BarnichEtAl:1995EinsteinWZ}. Theorem~5 therefore gives the following
sharper consequence.

\medskip
\noindent\textbf{Corollary (local anomaly freedom of the pure-Einstein spectator orbit).}
\emph{For the pure bosonic four-dimensional Einstein seed on a contractible spacetime without a
physical boundary, every regular $K=0$ disformal representative has
$H^{1,4}_{\rm loc}(s\mid d)=0$. If the quantum action principle applies and the regulator and
subtraction scheme preserve locality and the BV setup, every perturbative local breaking of the
quantum master equation is BRST exact and can be removed by local counterterms order by order in the
gravitational EFT expansion.}

This corollary discharges the \emph{local} gauge-anomaly hypothesis for that field content and domain.
It is not a statement about global gravitational anomalies, large diffeomorphisms, determinant-line
phases, zero modes, physical boundaries, singular orbits, or theories enlarged by chiral matter.

\subsection{Conditional all-order equivalence-theorem corollary}
\label{sec:all-order-orbit}

The invariance of physical observables under regular local field redefinitions is the equivalence
theorem, not a new theorem of the present work~\cite{Criado:2018sdb}. A recent analysis of a different
scalar duality also displays explicitly how a local differential-operator Jacobian reduces to
scaleless contact integrals in dimensional regularisation~\cite{deRham:2026dsd}. Here we specialize
that established logic to the complete disformal spectator BV orbit. The formal change-of-variables
statement gives an all-order perturbative corollary only under explicit hypotheses. Let
$F:q\mapsto Q$ be a regular, local and invertible field map on the chosen
patch, let $\Delta_F=\operatorname{Ber}(DF)$ be its full boson--fermion Berezinian, and transform the
sources, gauge fixing and boundary conditions together with the fields. Then
\begin{equation}
 Z_q[J_q]=\int\!Dq\,\Delta_F\,
 e^{iS_0[F(q)]+iJ_q\cdot q}
 =\int\!DQ\,e^{iS_0[Q]+iJ_Q\cdot Q}
 =Z_Q[J_Q].
 \label{eq:all-order-orbit}
\end{equation}
In dimensional regularisation the ultralocal contact part of $\ln\Delta_F$ is conventionally set to
zero; in another regulator it is a local measure contribution that must be retained. With compatible
BRST gauge fixing and boundary conditions, and with all seed, global and measure anomalies absent or
cancelled, the perturbative pole spectrum and physical observables are therefore identical at every
loop order along the regular orbit. The preceding corollary removes the local BV anomaly obstruction
for the pure four-dimensional Einstein spectator orbit under its stated assumptions, but does not
remove the global and boundary qualifications.
This is a disformal-orbit application of the equivalence theorem for the complete transformed
functional integral, not a claim that every
off-shell counterterm is equation-of-motion redundant and not an all-order proof of ghost-freedom for
arbitrary DHOST truncations.

\medskip
\noindent\textbf{Corollary (no frame-generated radiative protection).}
\emph{A regular disformal choice of field coordinates cannot supply a new Ward identity that protects
a propagating scalar built from the same field. Any genuine radiative protection must be inherited from
a seed symmetry (or from an independent mechanism such as a Galileon-type non-renormalisation theorem),
not from the frame transformation itself.}

\subsection{Relation to the known Horndeski and Einstein--scalar loop results}
\label{sec:loop-relation}

The loop calculation is deliberately organised as a frame-equivalence calculation, rather than as a
second derivation of the established Horndeski or Einstein--scalar divergences.  The constant-map
Einstein--Hilbert image is quartic Horndeski, while a regular $C(\phi,X),D(\phi,X)$ transformation is
an invertible change of field coordinates on its DHOST orbit.  We therefore take the known seed-frame
one-loop functionals as input, keep their complete equation-of-motion and measure terms, and pull the
correlated functional back with the same map that defines the classical theory.  The nontrivial result
is that the essential representative and the perturbative mode count can be followed explicitly on the
whole regular orbit; this is not a claim of a new universal Horndeski loop coefficient.  In particular,
the explicit canonical benchmark below is a consistency check of the orbit theorem and of the EFT
order-reduction procedure, while a complete determinant for arbitrary nonlinear $P(\widetilde X)$ is
outside the scope of this work.

For an Einstein seed the standard pure-gravity one-loop polynomial in harmonic background gauge is
\begin{equation}
 \mathcal B_g=\frac{1}{120}\tR^2+\frac{7}{20}\tR_{\mu\nu}\tR^{\mu\nu}.
 \label{eq:pure-gravity-pole}
\end{equation}
It is proportional to the leading Einstein equation and vanishes on Ricci-flat backgrounds, up to the
topological Euler term~\cite{tHooft:1974toh}. Its pole fixes a subtraction, not a finite
curvature-squared Wilson coefficient or a universal Stelle mass. Consequently no numerical massive
spin-2 pole follows from Eq.~\eqref{eq:pure-gravity-pole} alone.

\subsection{Complete canonical Einstein--scalar pole}

To test the propagating side of the obstruction theorem, we deliberately leave the $K=0$ spectator
orbit and choose the simplest shift-symmetric k-essence sector,
$K(\widetilde X)=-\widetilde X/2$~\cite{GarrigaMukhanov:1999,ArmendarizPicon:2000ah}. This canonical
choice breaks the local vertical factor to the global scalar shift and supplies one propagating scalar.
It is used because the complete coupled Einstein--scalar one-loop divergence is known, making it a
controlled benchmark rather than an assumption about general nonlinear k-essence. The seed is
\begin{equation}
 S_{\rm seed}=\int d^4x\sqrt{-\tg}
 \left[\frac{\MP^2}{2}\tR-\frac{1}{2}\tX\right].
 \label{eq:canonical-seed}
\end{equation}
The classic complete local one-loop result provides a decisive benchmark. In the rescaled field and
curvature conventions of Ref.~\cite{tHooft:1974toh}, its invariant polynomial is
\begin{equation}
 \boxed{
 \mathcal B_{g\phi}=\frac{1}{80}R^2+\frac{43}{120}R_{\mu\nu}R^{\mu\nu}
 +\frac{1}{2}Y^2-\frac{1}{12}RY+2(\Box\phi)^2,}
 \qquad Y=\nabla_\mu\phi\nabla^\mu\phi.
 \label{eq:full-Einstein-scalar-pole}
\end{equation}
The nonzero $2(\Box\phi)^2$ term shows that global shift symmetry does not forbid this off-shell
operator. Its coefficient can change under gauge and field redefinitions because the operator is
redundant; its presence nevertheless invalidates an operator-level assertion that gravitational loops
never generate it.

Define, in the same source convention,
\begin{equation}
 E_{\mu\nu}=R_{\mu\nu}-\frac{1}{2}g_{\mu\nu}R
 +\frac{1}{2}v_\mu v_\nu-\frac{1}{4}g_{\mu\nu}Y,
 \qquad E_\phi=\Box\phi,
 \qquad v_\mu=\nabla_\mu\phi.
 \label{eq:TV-EOM}
\end{equation}
There is an exact four-dimensional decomposition
\begin{equation}
 \boxed{
 \mathcal B_{g\phi}=\frac{203}{80}R^2+E^{\mu\nu}Q_{\mu\nu}+2E_\phi^2,}
 \label{eq:EOM-decomposition}
\end{equation}
where
\begin{equation}
 Q_{\mu\nu}=\frac{43}{120}R_{\mu\nu}+\frac{563}{240}g_{\mu\nu}R
 -\frac{43}{240}v_\mu v_\nu-\frac{523}{480}g_{\mu\nu}Y.
 \label{eq:Q-tensor}
\end{equation}
The $E^{\mu\nu}Q_{\mu\nu}+2E_\phi^2$ terms are removed at first order in the loop expansion by a local
perturbative field redefinition. They reappear in correlated form at higher orders and may not be
discarded when computing two-loop subdivergences. On the leading shell,
\begin{equation}
 \left.\mathcal B_{g\phi}\right|_{\rm EOM}
 =\frac{203}{80}R^2=\frac{203}{320}Y^2.
 \label{eq:on-shell-essential}
\end{equation}
Around a flat constant-scalar vacuum, this representative is a four-scalar contact interaction rather
than a new two-point pole. The on-shell coefficient $203/80$ appears explicitly in the gravity--scalar
literature~\cite{Benedetti:2009rx}. The original calculation here is the exact off-shell EOM ledger in
Eqs.~\eqref{eq:EOM-decomposition}--\eqref{eq:Q-tensor} and the
correlated regular-disformal pullback below.

\subsection{Restored normalization and orbit-wide disformal pullback}

Let $d=4-2\epsilon$. Restoring the canonical normalization in
Eq.~\eqref{eq:canonical-seed}, the magnitude of the essential local pole is
\begin{equation}
 \left|\Gamma_{\rm pole,ess}\right|
 =\frac{1}{16\pi^2\epsilon}\int d^4x\sqrt{-\tg}
 \frac{203}{80\MP^4}\tX^2.
 \label{eq:essential-pole}
\end{equation}
The overall sign depends on whether one quotes the divergent effective action or its subtraction
counterterm. The factor in Eq.~\eqref{eq:essential-pole} uses
$\epsilon=(4-d)/2$; choosing $4-d$ under the same prefactor doubles the displayed polynomial.

For the general map in four dimensions,
$\sqrt{-\tg}=C^{3/2}\sqrt{\W}\sqrt{-g}$ and $\tX=X/\W$ on the chosen Lorentzian branch. Therefore
\begin{equation}
 \boxed{
 \sqrt{-\tg}\,\tX^2
 =\sqrt{-g}\,\frac{C(\phi,X)^{3/2}X^2}
 {[C(\phi,X)-D(\phi,X)X]^{3/2}}.}
 \label{eq:essential-pullback}
\end{equation}
Therefore the essential one-loop pole has a first-derivative representative in the original frame,
\begin{equation}
 \delta P_{\rm ess}(\phi,X)=\frac{203}{80\MP^4}
 \frac{C(\phi,X)^{3/2}X^2}{[C(\phi,X)-D(\phi,X)X]^{3/2}},
 \label{eq:delta-P-essential}
\end{equation}
inside the pole convention of Eq.~\eqref{eq:essential-pole}. This is an orbit statement about the same
canonical Einstein--scalar theory in regular field coordinates, not a new determinant calculation for
an arbitrary nonlinear $P(X)$ seed. The complete off-shell pole also contains the correlated pullback
of the EOM terms in Eq.~\eqref{eq:EOM-decomposition}, together with the consistently transformed
measure discussed above; isolating and nonperturbatively resumming one term is not the one-loop EFT.

\subsection{Local stability domain}

After subtraction, parameterise the renormalised seed-frame scalar EFT through this order as
\begin{equation}
 P_{\rm ren}(\tX)=-\frac{1}{2}\tX+\alpha(\mu)\frac{\tX^2}{\MP^4},
 \qquad z=\frac{\alpha\tX}{\MP^4}.
 \label{eq:renormalized-P}
\end{equation}
The UV pole fixes the running contribution but not the finite matching value or sign of $\alpha$.
One finds
\begin{equation}
 P_{\tX}=-\frac{1}{2}+2z,
 \qquad
 P_{\tX}+2\tX P_{\tX\tX}=-\frac{1}{2}+6z,
 \label{eq:P-derivatives}
\end{equation}
so the local timelike no-ghost and no-gradient conditions hold for
\begin{equation}
 z<\frac{1}{12},
 \qquad
 c_s^2=\frac{1-4z}{1-12z}.
 \label{eq:canonical-health-domain}
\end{equation}
The controlled derivative expansion further requires $|z|\ll1$, fluctuation momenta, curvature and
background variation scales below the cutoff, and the regular branch $C>0$, $\W>0$, $\Jcal\neq0$.
If matter selects the original metric as its physical metric, compatibility of its causal cone must be
checked separately.

Equations~\eqref{eq:EOM-decomposition}--\eqref{eq:canonical-health-domain} support the following
limited quantum statement:
\begin{quote}
For the canonical Einstein--scalar local one-loop UV pole, the complete correlated pullback through
any regular $C(\phi,X),D(\phi,X)$ chart introduces no additional perturbative degree of freedom at
$O(\hbar)$ after EFT order reduction on the stated hyperbolic patch.
\end{quote}
We do not claim the absence of off-shell $(\Box\phi)^2$, equality of arbitrary off-shell 1PI
functionals, a complete finite or nonlocal effective action, a result for general nonlinear $P(X)$, or
an all-order theorem. The coupled graviton--scalar Hessian for general $P(X)$ has a nontrivial mixed
block and generically different graviton and scalar characteristic cones; its complete nonminimal
one-loop determinant remains open.

\section{Discussion: originality, mimetic gravity, and the healthy landscape}
\label{sec:discussion}

The central symmetry is a regular field-coordinate redundancy. Its Stueckelberg interpretation uses
the standard terminology~\cite{Ruegg:2003ps}. In DHOST variables, derivative-dependent
transformations can obscure the redundant direction completely.
The metric Jacobian of first-derivative disformal transformations was previously used to discuss
invertibility~\cite{Zumalacarregui:2013pma,DomenechEtAl:2015DerivativeMetric}. We therefore do not
claim that determinant in isolation. The original construction in this paper is the
closed-form spectator generator in Eq.~\eqref{eq:general-vertical} for the full
$C(\phi,X),D(\phi,X)$ class, the identification of the same $\Jcal$ in its denominator, and the
dynamical obstruction that closes the chain. This is not a relabelling of the known Jacobian: it
constructs the local gauge orbit explicitly and proves when it ceases to exist. Taken
together, these statements identify precisely how
the spectator fibre is embedded in a regular disformal chart and prove that the same scalar cannot
simultaneously be a locally shiftable spectator and carry a nonconstant $K(\phi,\tX)$ sector.

Proposition~3 separates a standard proof ingredient from the original result. For an invertible field
map, the transport of a pre-existing gauge identity is generic
~\cite{DomenechEtAl:2015DerivativeMetric,Takahashi:2017zgr}. What is derived here is the exact vertical representative,
its full regularity condition, and the variational-triviality criterion that classifies its existence
throughout the first-derivative disformal orbit. This is a specific covariant theorem for the full
regular $C(\phi,X),D(\phi,X)$ class, not a restatement of field-redefinition equivalence.

This also clarifies what it means for the gauge symmetry to constrain the conformal and disformal
data. In the constant-map ansatz the cancellation of derivative-parameter terms fixes the generator
coefficients in terms of $C_0,D_0$, while pulling back the Einstein seed locks the DHOST coefficient
functions to the single combination $\W$ through Eq.~\eqref{eq:gauge-selection-constant}. It does not
select numerical values of $C_0,D_0$. In the full regular class the factors are not forced to be
constant, because their $X$- and $\phi$-dependence is accompanied by the extra terms in the exact
generator~\eqref{eq:general-vertical}. What remains invariant under this enlargement is the seed mode
count: for the Einstein representative, two modes in the spectator theory and three after adding a
regular nonconstant same-field scalar sector, with no conclusion on the singular locus. The loop analysis follows the same logic: its coefficients
are imported from the known healthy seed and its original content is the correlated pullback and
the resulting orbit-wide EFT statement.

The relation to mimetic DHOST is complementary. Mimetic theories possess a known extra gauge symmetry
because the auxiliary-to-physical metric map is non-invertible
~\cite{Langlois_2019,Domenech:2023ryc,Gorji:2026dmc}. In that case the
Jacobian has a null vector, the map may change the degree-of-freedom content, and stability is a separate
issue. Here $\Jcal\neq0$: the finite vertical group is obtained by conjugating a spectator translation
with an invertible map. Our originality claim is therefore the explicit regular-disformal spectator
mechanism and its obstruction, not the existence of gauge symmetry somewhere in the broader DHOST
landscape.

Nor is quartic Horndeski the only healthy scalar--tensor theory. The constant Einstein image happens to
lie in quartic Horndeski. General invertible $C(\phi,X),D(\phi,X)$ maps organise wider class-Ia DHOST
families, and the full
Horndeski, beyond-Horndeski, quadratic and cubic DHOST landscape contains many other structurally
degenerate theories~\cite{BenAchour:2016bwh,BenAchour:2016fzp,Kobayashi_2019}. Their actual health also
depends on background kinetic signs, hyperbolicity, strong coupling and matter coupling. Our theorem
classifies a symmetry inherited from a spectator seed; it does not classify every healthy theory.

The quantum result likewise separates symmetry from EFT equivalence. In the spectator sector
Theorem~5 proves that the full antifield-dependent local BV cohomology reduces to that
of the seed complex: counterterms, gauge deformations and anomaly candidates acquire no independent
spectator class. For the pure four-dimensional Einstein seed on a contractible spacetime, the known
seed cohomology then removes the perturbative local anomaly obstruction, while the single-field
threefold (and multi-field reduced) Jacobian tracks the regular coordinate and measure factors. Under the explicit
hypotheses of Sec.~\ref{sec:all-order-orbit}, this becomes a conditional all-order orbit-equivalence
corollary for physical observables, not a new equivalence theorem and not a statement that all
off-shell counterterms are redundant. In the
dynamical canonical sector the local gauge factor is absent. Its one-loop statement follows instead
from the explicit EOM reduction of the complete pole and its orbit-wide regular pullback. A global shift
allows higher-derivative scalar operators; it does not provide the missing local Ward identity. Enhanced
global symmetries such as Galileon or
weakly broken Galileon symmetry can lead to other non-renormalisation statements
~\cite{de_Rham_2013,Santoni_2018,Heisenberg_2020}, but they are different mechanisms.

\section{Conclusions}
\label{sec:conclusions}

An AI-assisted search suggested a local transformation in a constant-disformal scalar--tensor theory.
Detailed analysis did more than verify the candidate: it changed its interpretation and enlarged its
scope. The transformation is a field-dependent diffeomorphism plus a local translation of a spectator
scalar. Pulling the transparent translation through the general regular
$C(\phi,X),D(\phi,X)$ map gives the exact generator in Eq.~\eqref{eq:general-vertical}. Its denominator
is $\Jcal$, which also controls the kinetic inverse and the determinant of the metric-coordinate map.

This symmetry explains why a two-mode metric seed such as Einstein gravity, rewritten with a redundant
scalar, still has two physical modes and correlated counterterms. It does not protect a dynamical scalar. For any nonconstant
$K(\phi,\tX)$, the local vertical factor is obstructed off shell. A global shift remains only in the
shift-symmetric subcase; ordinary diffeomorphisms remain throughout. The $\phi$-dependent
cuscuton-type square-root branch is principal-degenerate but is not this gauge symmetry.

For several scalars the exact criterion is variational triviality. When $n<d$, this again forces
$K$ to be constant on a generic open jet branch. When $n\geq d$, determinant/minor null Lagrangians
are nonconstant topological exceptions. They preserve the local spectator group precisely because
they add no bulk scalar equation; they are neither propagating k-essence nor a singular mimetic
mechanism.

Our explicit spectator and BV construction, combined with the standard regular-orbit lemma,
makes the quantum interpretation precise: an invertible local DHOST map cannot itself create an
independent pole or an Ostrogradsky pair when its transformed measure is included. With the
qualifications in Sec.~\ref{sec:all-order-orbit}, the physical spectrum is orbit-equivalent to all
perturbative loop orders; any quantum ghost must instead come from the seed
dynamics, radiative operators, the EFT truncation, a measure/anomaly contribution, or a singular map.
This is the strongest model-independent route back to the title, and it is deliberately an
orbit-equivalence statement, not an assertion that every DHOST theory is healthy.

For the canonical propagating theory, the complete one-loop pole contains an off-shell
$(\Box\phi)^2$ term. Its exact EOM decomposition leaves one essential $\tX^2$ representative, whose
pullback is the first-derivative counterterm Eq.~\eqref{eq:delta-P-essential} throughout the regular
disformal orbit. Consequently the same canonical theory acquires no additional perturbative mode at
$O(\hbar)$ merely by being expressed in any of these regular charts, after order reduction within the
declared EFT and hyperbolicity domain. No universal Stelle mass, finite Wilson coefficient,
general-$P(X)$ determinant or all-order ghost-freedom theorem follows.

The title is therefore understood in a precise perturbative sense. Regular-disformal DHOST
representatives can be free of additional quantum modes in a controlled EFT expansion, while the local
symmetry found by the search protects only the spectator sector. Theorem~5 sharpens this statement:
the complete spectator BV quartet supplies no new local counterterm, gauge-deformation or anomaly
class beyond the seed complex. For the pure Einstein seed the local anomaly group vanishes under the
stated four-dimensional contractible-domain assumptions, although global and boundary questions
remain. The principal original outcome is the
single-field threefold (and multi-field reduced) Jacobian correspondence linking the explicit spectator
generator to regular field coordinates, the dimension-dependent topological boundary of the
multi-field obstruction, and the explicit disformal BV-quartet reduction, together with an orbit-wide
one-loop demonstration that the canonical theory remains perturbatively three-mode on the stated
domain. The conditional all-order statement is a consequence of the established equivalence theorem
under the hypotheses stated in Sec.~\ref{sec:all-order-orbit}; its value here is to complete, rather
than replace, the original spectator construction.

\acknowledgments
We thank the developers of \emph{Denario} for making the tool available. This work was supported by a
grant from the Simons Foundation (00017375, RJ). Funding for the work of RJ was partially provided by
project PID2022-141125NB-I00, and the ``Center of Excellence Maria de Maeztu 2025-2029'' award to the
ICCUB funded by grant CEX2024-001451-M from AEI/10.13039/501100011033.

\appendix

\section{Constant-disformal projection}
\label{app:projection}

Let $H_{\mu\nu}=\nabla_\mu\nabla_\nu\phi$, $B=\Box\phi$,
$\Psi_\mu=\phi^\rho H_{\rho\mu}$ and $\Phi=\phi^\mu\phi^\nu H_{\mu\nu}$.
For Eq.~\eqref{eq:constant-map}, metric compatibility and symmetry of the scalar Hessian give
Eq.~\eqref{eq:connection-difference}. With $\kappa=D_0/\W$ and
$\kappa_X=\kappa^2$, direct curvature contraction yields
\begin{align}
 C_0\tR={}&R+2\kappa R_{\mu\nu}\phi^\mu\phi^\nu
 +\kappa(H_{\mu\nu}H^{\mu\nu}-B^2)
 -2\kappa_X(\Phi B-\Psi_\mu\Psi^\mu).
 \label{eq:raw-Ricci-projection}
\end{align}
Multiplying by the exact measure gives a pointwise raw density with
\begin{equation}
 \bar A_1=\mu D_0\sqrt{\frac{C_0}{\W}},
 \qquad
 \bar A_3=-2\mu D_0^2\frac{\sqrt{C_0}}{\W^{3/2}}.
 \label{eq:raw-A}
\end{equation}
The Ricci identity
\begin{equation}
 R_{\mu\nu}\phi^\mu\phi^\nu
 =B^2-H_{\mu\nu}H^{\mu\nu}+\nabla_\mu
 (\phi^\nu H^\mu{}_{\nu}-\phi^\mu B)
 \label{eq:Ricci-reduction}
\end{equation}
must be integrated with the $X$-dependent $\bar A_1$. The resulting
$\bar A_{1,X}$ contribution cancels the raw $\bar A_3$ structures because
$4\bar A_{1,X}=-\bar A_3$. This leaves Eq.~\eqref{eq:projected-coefficients}.
The calculation distinguishes the raw pointwise identity from equality of actions modulo a boundary
term.

\section{Gauge algebra, BRST data, and canonical count}
\label{app:gauge}

For independent field-independent parameters the transparent algebra is
Eq.~\eqref{eq:semidirect-algebra}. Introduce the odd diffeomorphism ghost $c^\mu$ and vertical ghost
$\rho$. With a left BRST differential,
\begin{equation}
 s\tg_{\mu\nu}=\mathcal L_c\tg_{\mu\nu},
 \quad
 s\phi=c^\mu\phi_\mu+\rho,
 \quad
 sc^\mu=c^\nu\partial_\nu c^\mu,
 \quad
 s\rho=c^\nu\partial_\nu\rho.
 \label{eq:BRST}
\end{equation}
These transformations are nilpotent off shell. For the constant map their pullback is
\begin{equation}
 sg_{\mu\nu}=\mathcal L_cg_{\mu\nu}
 +\frac{D_0}{C_0}(\phi_\mu\partial_\nu\rho+\phi_\nu\partial_\mu\rho).
 \label{eq:BRST-pullback}
\end{equation}
The minimal BV master action is
\begin{align}
 S_{\rm min}=S[\tg]+\int d^dx\big[&\tg^{*\mu\nu}\mathcal L_c\tg_{\mu\nu}
 +\phi^*(c^\mu\phi_\mu+\rho)\nonumber\\
 &-c^*_\mu c^\nu\partial_\nu c^\mu-\rho^*c^\nu\partial_\nu\rho\big],
 \label{eq:BV-master}
\end{align}
which satisfies the classical master equation under the standard boundary assumptions. By itself this
is a classical algebraic statement; the anticanonical lift, full quartet reduction and scoped local
anomaly corollary are established separately in Sec.~\ref{sec:spectator-cohomology}.

For the Einstein--Hilbert seed, in ADM transparent variables the $K=0$ scalar momentum vanishes and generates the vertical shift.
The lapse and shift enforce the usual four diffeomorphism constraints. Counting only
$(\widetilde h_{ij},\phi)$ gives
\begin{equation}
 N_{\rm dof}=\frac{1}{2}\left[2(6+1)-2(4+1)\right]=2.
 \label{eq:ADM-count-K0}
\end{equation}
For a regular, nondegenerate kinetic $K(\phi,\tX)$, $\pi_\phi$ is no longer constrained and
\begin{equation}
 N_{\rm dof}=\frac{1}{2}\left[2(6+1)-2(4)\right]=3.
 \label{eq:ADM-count-K}
\end{equation}
For a different metric seed, the corresponding statement is $N_{\rm phys}=N_{\rm seed}$ or
$N_{\rm seed}+1$, respectively. The displayed count assumes no further accidental gauge symmetry or
kinetic degeneracy.

\section{One-loop coefficient and EOM ledger}
\label{app:oneloop}

The regulator convention in Eq.~\eqref{eq:essential-pole} is
$\epsilon=(4-d)/2$. If instead $\epsilon_1=4-d$, then
$1/\epsilon=2/\epsilon_1$. Keeping a universal-looking
$1/(16\pi^2\epsilon_i)$ prefactor therefore requires doubling all polynomial coefficients when moving
from $\epsilon$ to $\epsilon_1$.

The contractions needed to verify Eq.~\eqref{eq:EOM-decomposition} are
\begin{align}
 E^{\mu\nu}R_{\mu\nu}
 &=R_{\mu\nu}^2-\frac{1}{2}R^2+\frac{1}{2}R_{\mu\nu}v^\mu v^\nu-\frac{1}{4}RY,
 \nonumber\\
 ER&=-R^2-\frac{1}{2}RY,
 \nonumber\\
 E^{\mu\nu}v_\mu v_\nu
 &=R_{\mu\nu}v^\mu v^\nu-\frac{1}{2}RY+\frac{1}{4}Y^2,
 \nonumber\\
 EY&=-RY-\frac{1}{2}Y^2.
 \label{eq:EOM-contractions}
\end{align}
Substitution of Eq.~\eqref{eq:Q-tensor} cancels the mixed
$R_{\mu\nu}v^\mu v^\nu$ invariant and reproduces every coefficient in
Eq.~\eqref{eq:full-Einstein-scalar-pole}. The perturbative change
\begin{equation}
 g_{\mu\nu}\mapsto g_{\mu\nu}-\lambda Q_{\mu\nu},
 \qquad
 \phi\mapsto\phi-2\lambda E_\phi,
 \qquad \lambda=O(\hbar),
 \label{eq:EOM-field-redefinition}
\end{equation}
removes the EOM terms to first order, with the overall sign reversed if $\lambda$ denotes the already
subtracted counterterm rather than the divergence. Its inverse is perturbatively local; treating it as
an exact finite higher-derivative transformation is neither necessary nor generally valid.

The source uses a unit gravitational coupling and a rescaled scalar. With
$\kappa^2=2/\MP^2$ and $Y_{\rm TV}=\kappa^2\tX$, the essential source
polynomial becomes
\begin{equation}
 \frac{203}{320}Y_{\rm TV}^2=\frac{203}{80\MP^4}\tX^2,
 \label{eq:normalization-restoration}
\end{equation}
which establishes Eq.~\eqref{eq:essential-pole}.

\section{Reproducibility and scope of the checks}
\label{app:reproducibility}

The accompanying validation harness separates results into hard-failing gates. It checks:
\begin{enumerate}
 \item conventions, branches, inverse metrics and functional Jacobians;
 \item the constant projection through independent covariant and component calculations;
 \item the complete Diff-plus-shift algebra, nilpotency and classical BV master equation;
 \item the formal $K=0$ measure reduction, the anticanonical spectator-ghost coordinate, the split
 master action and the full antifield-dependent BV quartet cohomology;
 \item the pure-gravity and complete canonical Einstein--scalar one-loop coefficient ledgers;
 \item the exact EOM decomposition and orbit-wide canonical essential pullback;
 \item the acoustic determinant and local hyperbolicity domain;
 \item the $K(\phi,\tX)$ Noether obstruction and $\phi$-dependent cuscuton branch;
 \item the general $C(\phi,X),D(\phi,X)$ vertical generator on non-diagonal component metrics;
 \item the symmetric-metric determinant, including $\det_{\rm sym}\mathbb A=C^9\Jcal$ in four
 dimensions;
 \item the exact finite orbit in a nonconstant example;
 \item the multi-field reduced Jacobian, its $n=2$ symbolic/Mathematica check, the corrected
 $[\det W]^{n+1}$ kinetic factor, noncommuting matrix ordering, and the dimension-saturating
 null-Lagrangian exception;
 \item the four-dimensional pure-Einstein local anomaly corollary and the conditional all-order
 change-of-variables statement with its global, boundary, gauge-fixing and regulator exclusions.
\end{enumerate}
The full suite contains deliberate mutations of signs, normalisations and omitted terms; each mutation
must produce a nonzero residual. These checks validate the displayed local identities. They do not
replace the unperformed complete one-loop calculation for nonlinear $P(X)$, a global analysis of
hyperbolicity, a BV--BFV boundary analysis, or a global-anomaly calculation.

\providecommand{\href}[2]{#2}\begingroup\raggedright\endgroup

\end{document}